\documentclass{article}
\usepackage{fleqn,mprocl,amssymb}
\usepackage{graphicx}

\newcommand{\AmS}{{\protect\the\textfont2
  A\kern-.1667em\lower.5ex\hbox{M}\kern-.125emS}}
\def\be{\begin{equation}}
\def\ee{\end{equation}}
\def\bea{\begin{eqnarray}}
\def\eea{\end{eqnarray}}

\begin{document}

\title{ELECTRON-PHONON COUPLING ORIGIN OF THE RESISTIVITY\\
IN YNI$_2$B$_2$C SINGLE CRYSTALS}

\author{R. S. GONNELLI, A. MORELLO, G. A. UMMARINO}

\address{INFM - Dipartimento di Fisica, Politecnico di Torino,
c.so Duca degli Abruzzi, 24 - 10129 Torino, Italy}

\author{V. A. STEPANOV}

\address{P.N. Lebedev Physical Institute, Russian Academy of Sciences, \\
Leninski Pr. 53, Moscow, Russia}

\author{G. BEHR, G. GRAW, S. V. SHULGA, S. -L. DRECHSLER}

\address{Instit\"{u}t f\"{u}r Festk\"{o}rper- und Werkstofforschung Dresden, \\
Postfach 270016, D-01171 Dresden, Germany}


\maketitle \abstracts{Resistivity measurements from 4.2~K up to
300~K were made on YNi$_2$B$_2$C single crystals with
$T_{\mathrm{c}}$~=15.5~K. The resulting $\rho(T)$ curve shows a
perfect Bloch-Gr\"{u}neisen (BG) behavior, with a very small
residual resistivity which indicates the low impurity content and
the high cristallographic quality of the samples. The value
$\lambda_{\mathrm{tr}}$~=0.53 for the transport electron-phonon
coupling constant was obtained by using the high-temperature
constant value of d$\rho$/d$T$ and the plasma frequency reported
in literature. The BG expression for the phononic part of the
resistivity $\rho_{\mathrm{ph}}(T)$ was then used to fit the data
in the whole temperature range, by approximating
$\alpha^{2}_{\mathrm{tr}}F(\Omega)$ with the experimental phonon
spectral density $G(\Omega)$ multiplied by a two-step weighting
function to be determined by the fit. The resulting fitting curve
perfectly agrees with the experimental points. We also solved the
real-axis Eliashberg equations in both $s$- and $d$-wave
symmetries under the approximation $\alpha^{2}F(\Omega)\approx
\alpha^{2}_{\mathrm{tr}}F(\Omega)$. We found that the value of
$\lambda_{\mathrm{tr}}$ here determined in single-band
approximation is quite compatible with $T_{\mathrm{c}}$ and the
gap $\Delta$ experimentally observed. Finally, we calculated the
normalized tunneling conductance, whose comparison with
break-junction tunnel data gives indication of the possible
$s$-wave symmetry for the order parameter in YNi$_2$B$_2$C.}

\section{Introduction}
In the past seven years, a great interest has been aroused by the
discovery of the family of superconducting quaternary rare-earth
borocarbide intermetallic compounds RNi$_2$B$_2$C (R~=~rare
earth), of which YNi$_2$B$_2$C is one of the most studied
non-magnetic members. In some way similarly to high-$T_{\rm c}$
cuprates, the borocarbides have a layered crystal structure, even
if band-structure calculations have shown the presence of a
three-dimensional electronic structure (see Ref.~[1] for a short
overview of selected properties). Initially, a conventional BCS
description of the superconducting properties in these materials
has been supported by many of the experimental and theoretical
results. More recently some particular features of the electronic
specific heat (as a function of temperature and magnetic field)
and of the upper critical field in the non-magnetic borocarbides
LuNi$_2$B$_2$C and YNi$_2$B$_2$C have been interpreted as signs of
an unconventional pairing, possibly of $d$-wave type \cite{ref0}.
On the other hand, Shulga \emph{et al.} have recently explained
the positive curvature of the upper critical field of
YNi$_2$B$_2$C as a function of the temperature near
$T_\mathrm{c}$, its magnitude and shape in the framework of
$s$-wave Migdal-Eliashberg theory by considering the presence of
two bands, one of them being more deeply involved in the transport
properties of the compound~\cite{ref2}.
\begin{figure}[t]
\hspace{1cm}\includegraphics[keepaspectratio,width=9cm]{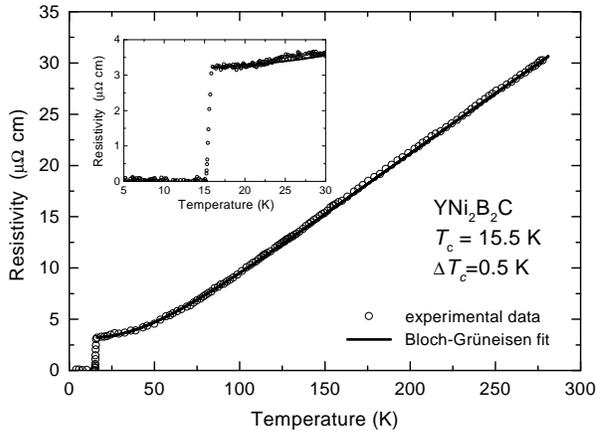}
\vspace{-3mm}\caption{A few points of the typical resistivity of
one of our YNi$_2$B$_2$C single crystals (open circles) and the
fit using the Bloch-Gr\"{u}neisen model (solid line). The inset
shows the low temperature part of the resistivity and the fit with
all the measured points. \label{fig:radish}}
\end{figure}

In the present work we demonstrate the complete agreement of the
experimental resistivity data obtained in YNi$_2$B$_2$C with the
predictions of the classic single-band theory for the
electron-phonon (e-p) interaction (Bloch-Gr\"{u}neisen theory). We
obtain a value of the transport e-p coupling constant that well
agrees with previous experimental and theoretical results. The
transport e-p spectral function
$\alpha^{2}_{\mathrm{tr}}F(\Omega)$ is also obtained by the fit of
the experimental data in the whole temperature range and used for
the calculation of the normalized tunneling conductance by
directly solving the real-axis Eliashberg equations.

\section{Experiment}
High-quality YNi$_2$B$_2$C single crystals were grown by using the
rf - zone melting process \cite{ref3}. The critical temperature of
the crystals, measured by AC susceptibility and perfectly
confirmed by resistivity measurements, is $T_{\rm c} = 15.5\,$K
with $\Delta T_{\rm c}(10-90\%) = 0.5\,$K. The imaginary part of
the susceptibility shows a single, very high and narrow peak at
$T \approx T_\mathrm{c}$ which confirms the purity and
crystallographic quality of the samples.

We measured the resistivity of these crystals as a function of the
temperature, on cooling from 300 K down to 4.2 K, by using the AC
version of the standard four-probe technique. The current leads
were directly soldered to the opposite sides of the samples, which
have parallelepipedal shape. The voltage leads, made from very
thin gold wires, were glued with a conducting paste to the surface
of the crystals. We improved the sensitivity of the measurement by
injecting in the crystals an AC current of typically 10 mA at 133
and detecting the voltage by the standard lock-in technique. Due
to a very slow cooling-down procedure, we were able to collect
nearly three thousand resistivity values for every curve between
4.2 and 300~K.

In Figure 1 the resistivity $\rho(T)$ of one of the YNi$_2$B$_2$C
crystals is shown (open circles). For clarity, only a reduced
number of points is reported. In the inset of the same figure we
show an enlargement of the low-temperature part of the resistivity
curve that contains all the measured points at $T < 30$ ~K (open
circles).

The resistivity of Fig.~1 shows a {\it perfect}
Bloch-Gr\"{u}neisen (BG) behavior that has already been observed
in previous experiments \cite{ref4}. The small residual value of
the resistivity $\rho(0)$=~3 $\mu\Omega \cdot $cm and its
perfectly linear high-temperature part (with a slope
$\mathrm{d}\rho /\mathrm{d}T$ =~0.12 $\mu \Omega \cdot $cm/K)
indicate the high quality and low impurity content of the samples.
Quite similar results were obtained in various YNi$_2$B$_2$C
samples.

\begin{figure}[t]
\hspace{1cm}\includegraphics[keepaspectratio,width=9cm]{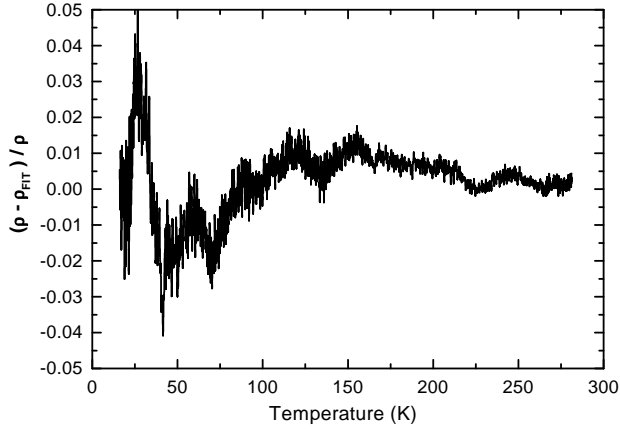}
\vspace{-6mm}\caption{The relative deviations
\mbox{$(\rho\!-\!\rho _{\rm{FIT}})\!/\!\rho$} of the experimental
data from the fit.}
\end{figure}

\section{Discussion}
The resistivity of a normal Fermi-liquid metal follows the well
known Matthiessen's rule: $\rho (T)$=$\rho _0$+$\rho
_{\rm{ph}}(T)$, where $\rho _0$ and $\rho _{\rm{ph}}(T)$ are the
residual and the phonon resistivities, respectively.

According to the BG theory, the high-temperature part of the
$\rho_{\rm{ph}}(T)$ shows a linear behavior given by the
following expression:

\[
\rho_{\rm{ph}}(T)=(2\pi \varepsilon _{0}k_{\rm{B}}/\hbar \omega
_{\rm{p}}^{2})\lambda _{\rm{tr}}T
\]
where $\omega_{\rm{p}}$ is the plasma frequency, $\lambda_{\rm{tr}}%
$=$2\int_0^\infty [\alpha_{\rm{tr}}^2 F(\Omega)/\Omega ]d\Omega$
is the transport \mbox{e-p} coupling constant, and $\alpha_{\rm{tr}}^2%
F(\Omega )$ is the transport e-p spectral function. Once known the
experimental value of the plasma energy $\hbar\omega_\mathrm{p}$
and the temperature coefficient of the linear part of the
resistivity at $T\!\!>\!\!$ 100~K (see Fig.~1), we can determine
the transport coupling constant $\lambda_{\rm{tr}}\!=\!(\hbar
\omega _{\rm{p}}^{2}/2\pi \varepsilon
_{0}k_{\rm{B}})\mathrm{d}\rho_{\mathrm{ph}}/\mathrm{d}T
\!=\!(\hbar \omega _{\rm{p}}^{2}/2\pi \varepsilon
_{0}k_{\rm{B}})\mathrm{d}\rho/\mathrm{d}T$. The values
$\omega_{\rm{p}}\!=\!$ 4.25 eV determined in Ref.~[6] by
reflectance and EELS measurements and $\mathrm{d}\rho
/\mathrm{d}T\!=\!$~0.12 $\mu \Omega \cdot $cm/K from Fig.~1 lead
to the result $\lambda_{\rm{tr}}\!=\!$ 0.53.

Full information on the \mbox{e-p} coupling in YNi$_2$B$_2$C can
be obtained by fitting the resistivity of Fig.~1 in the whole
temperature range. Again following the standard BG theory, we use
the most general expression for $\rho_{\mathrm{ph}}(T)$, given by:
\[
\rho_{\rm{ph}}(T)=(4\pi \varepsilon _{0}k_{\rm{B}} T/\hbar \omega
_{\rm{p}}^{2})\int_{0}^{\Omega _{\max }}[\alpha
_{\rm{tr}}^{2}F(\Omega )/\Omega ] \cdot[\hbar
\Omega/2k_{\rm{B}}T\sinh (\hbar \Omega /2k_{\rm{B}}T)]^{2}d\Omega.
\]

Our goal is to determine the function $\alpha_{\rm{tr}}^2
F(\Omega)$ from the fit. Actually, as a first approximation, for
the $\alpha_{\rm{tr}}^2 F(\Omega)$ we use the phonon spectral
density $G(\Omega)$ obtained in inelastic neutron scattering
experiments \cite{ref6} multiplied by a two-step weighting
function, whose constant values for $\Omega\! <\! 37.5$ meV and
$37.5\!<\! \Omega\! <\!70$ meV are to be determined by the fit.
These energy ranges are chosen because they correspond to the two
most-distinguishable structures of the $G(\Omega)$ which shows
peaks at about 20 and 50 meV and a value close to zero just at
37.5 meV. We neglected in the fit the presence of high energy
phonons at about 100 meV.

The results of the fit are shown as a solid line in Fig.~1. The
experimental $\rho (T)$ is \emph{perfectly} fitted by the
theoretical BG curve in the whole temperature range. Fig.~2 shows
the relative deviations of the experimental curve from the fit
$(\rho -\rho _{\rm{FIT}})/\rho$ as function of temperature. They
never exceed $\pm 5\%$.

Fig.~3 shows the spectral function $\alpha_{\rm{tr}}^2 F(\Omega )$
obtained as the product of the $G(\Omega)$ by the two-step
weighting function $\alpha_{\rm{tr}}^2$ determined from the fit.
The latter function is shown in the inset of Fig.~3.  Of course,
the $\lambda_{\rm{tr}}$ calculated from the $\alpha_{\rm{tr}}^2
F(\Omega )$ perfectly coincides with the value previously obtained
from the linear part of $\rho (T)$.

\begin{figure}[t]
\hspace{1cm}\includegraphics[keepaspectratio,width=9cm]{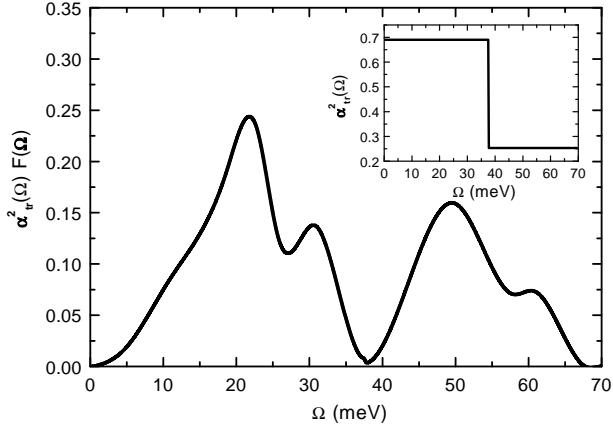}
\vspace{-5mm}\caption{The electron-phonon spectral function
determined from the fit of the resistivity of Fig.~1. In the
inset the step-like e-p coupling function obtained from the fit
is shown.}
\end{figure}

It is well known that, to a first approximation,
$\lambda_{\rm{tr}} \approx \lambda$, where $\lambda$ is the e-p
coupling factor involved in the BCS coupling of the Cooper's
pairs. In the hypothesis of an e-p coupling origin of the
superconductivity in YNi$_2$B$_2$C we calculate the quasiparticle
density of states in this compound  in both $s$- and $d$-wave
symmetry by solving in direct way the real-axis Eliashberg
equations for the strong e-p coupling~\cite{ref6b} and using as
$\alpha_{\rm{tr}}^2 F(\Omega)$ the function shown in Fig. 3.

\begin{figure}[t]
\hspace{1cm}\includegraphics[keepaspectratio,width=9cm]{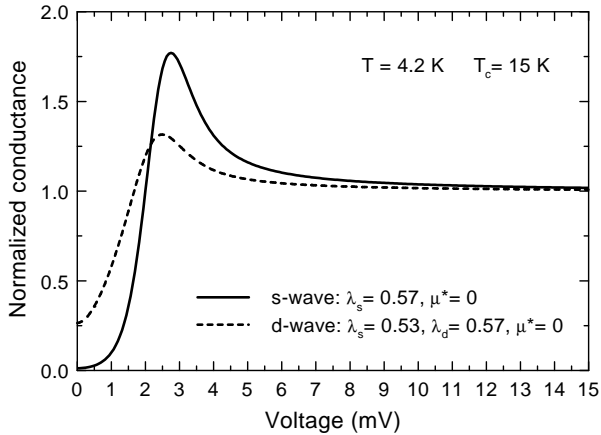}
\vspace{-6mm}\caption{$s$- and $d$-wave SIN tunneling conductance
of YNi$_2$B$_2$C calculated by direct solution of the Eliashberg
equations using the e-p spectral function shown in Fig.~3 and the
parameters shown in the legend.}
\end{figure}

Fig.~4 shows the SIN tunneling conductances in $s$- and $d$-wave
symmetry calculated at 4.2~K from the density of states (solid and
dash lines, respectively). From the solution of the Eliashberg
equations we obtained both the correct $T_{\rm c}$ and the
superconducting gap $\Delta\approx$ 2 meV observed in tunneling
experiments \cite{ref7,ref8,ref9}, by using a value
$\lambda$=~0.57 slightly greater than $\lambda_{\rm{tr}}$. This
fact is consistent with the conventional relation between
$\lambda$ and $\lambda_{\rm{tr}}$. From the curves of Fig.~4 and
the tunneling experimental data in STM configuration
\cite{ref8,ref9} it is very difficult to determine the possible
symmetry of the order parameter in YNi$_2$B$_2$C due to the
appreciable broadening of the SIN data.

\begin{figure}[thb]
\hspace{1cm}\includegraphics[keepaspectratio,width=9cm]{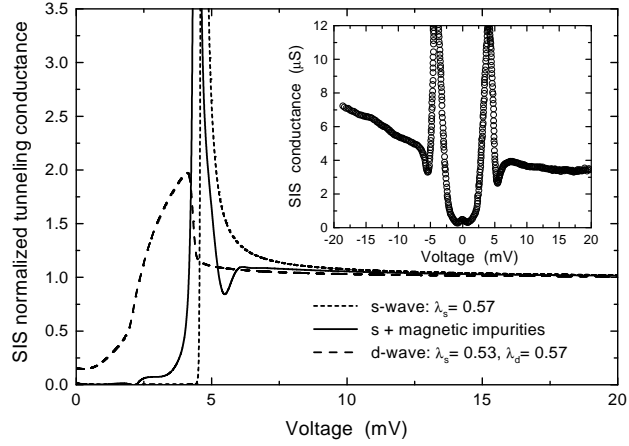}
\vspace{-3mm}\caption{$s$- and $d$-wave SIS tunneling conductance
of YNi$_2$B$_2$C (dot and dash line, respectively) calculated by
direct solution of the Eliashberg equations using the e-p spectral
function shown in Fig.~3 and the parameters indicated in the
legend. The solid line represents the $s$-wave SIS conductance for
$\lambda$=~0.57 in presence of a small amount of magnetic
impurities corresponding to the parameters $C_{\mathrm{0}}$=4 and
$\Gamma$=0.8 of the model of Ref.~[12]. The inset shows the
break-junction tunneling conductance at 4.4~K from Ref.~[9].}
\end{figure}

A more clear indication can be obtained by the comparison of the
calculated SIS tunneling conductance with the break-junction
experimental data present in literature. This comparison is shown
in Fig. 5 where the theoretical SIS curves at 4 K determined from
the solution of Eliashberg equation for pure $d$-wave (dash line),
pure $s$-wave (dot line) and $s$-wave plus magnetic impurities in
quasi non-unitary limit \cite{ref10} (solid line) are presented
together with the break-junction data of Ekino {\it et al}
\cite{ref7} (inset). The $s$-wave curve in presence of a small
amount of magnetic impurities (for details see the caption of the
figure and Ref.~[12]) reproduces all the main features of the
experimental data including the well pronounced dip at about twice
the energy of the gap. On the other hand, the $d$-wave tunneling
conductance is quite different from the experimental curve. These
results suggest a possible $s$-wave symmetry (or, at least, a
dominant $s$-wave component) in YNi$_2$B$_2$C and give evidence
for the essential role played by the electron-phonon coupling in
the pairing mechanism in this compound.
\section{Conclusions}
We have demonstrated that the resistivity of YNi$_2$B$_2$C has a
temperature dependence \emph{perfectly} described by the standard
BG theory for the e-p coupling in conventional metals. The value
of $\lambda_{\rm{tr}}$ here determined from the resistivity is
representative of an intermediate e-p coupling strength and is
consistent with the value used in Ref.~[3] for the discussion of
the superconducting properties of YNi$_2$B$_2$C in the framework
of the isotropic single-band model. The transport e-p spectral
function $\alpha_{\rm{tr}}^2 F(\Omega)$ determined from the fit of
the resistivity was used, to a first approximation, in the direct
solution of the Eliashberg equations both in $s$- and $d$-wave
symmetry. The resulting $T_{\mathrm{c}}$ and $\Delta$ are quite in
agreement with the experimental data, while the comparison of the
calculated SIS tunneling conductance with the break-junction data
supports a conventional $s$-wave symmetry for the order parameter
in YNi$_2$B$_2$C.

\end{document}